\documentclass[12pt]{spieman}  
\usepackage{amsmath,amsfonts,amssymb}
\usepackage{graphicx}
\usepackage{setspace}
\usepackage{tocloft}
\usepackage{lineno}
\usepackage{enumerate}
\title{Simulations of the {Spectral Resolving Power} of a Compact Space-Borne Immersion-Echelle Spectrometer Using Mid-Infrared Wave Tracing}

\author[a]{Satoshi Itoh}
\author[a]{Daisuke Ishihara}
\author[a]{Takehiko Wada}
\author[a]{Takao Nakagawa}
\author[b]{Shinki Oyabu}
\author[c]{Hidehiro Kaneda}
\author[d]{Yasuhiro Hirahara}
\author[ ]{the SMI consortium}

\affil[a]{Department of Space Astronomy and Astrophysics,
Institute of Space and Astronautical Science,
Japan Aerospace Exploration Agency,   3-1-1 Yoshinodai Chuo-ku, Sagamihara, Japan, 252-5210}
\affil[b]{Institute of Liberal Arts and Sciences, Tokushima University, 1-1 Minami Jousanjima-machi, Tokushima, Japan, 770-8502}
\affil[c]{Graduate School of Science, Nagoya University, Furo-cho Chikusa-ku, Nagoya, Japan, 464-8601}
\affil[d]{Graduate School of Environmental Studies, Nagoya University, Furo-cho Chikusa-ku, Nagoya, Japan, 464-8601}

\cftpagenumbersoff{figure}
\cftpagenumbersoff{table} 
\begin{document} 
\maketitle

\begin{abstract}
We performed wave-optics-based numerical simulations at mid-infrared wavelengths to investigate how the presence or absence of entrance slits and optical aberrations affect the {spectral resolving power} $R$ of a compact, {high-spectral-resolving-power} spectrometer containing an immersion-echelle grating.
We tested three cases of telescope aberration (aberration-free, astigmatism and spherical aberration), assuming the aberration budget of the Space Infrared Telescope for Cosmology and Astrophysics (SPICA), which has a 20-$\mathrm{\mu m}$-wavelength diffraction limit.  
In cases with a slit, we found that the value of $R$ at around 10--20 $\mathrm{\mu m}$ is approximately independent of the assumed aberrations, which is significantly different from the prediction of geometrical optics. 
Our results also indicate that diffraction from the slit improves $R$ by enlarging the effective illuminated area on the grating window and that this improvement decreases at short wavelengths.   
For the slit-less cases, we found that the impact of aberrations on $R$ can be roughly estimated using the Strehl ratio. 
\end{abstract}


{\noindent \footnotesize\textbf{*}Satoshi Itoh,  \linkable{sitoh@ir.isas.jaxa.jp} }


\section{Introduction}
Space-borne spectroscopy with a {spectral resolving power} higher than tens of thousands at mid-infrared (MIR) wavelengths is an unexplored region of astronomy that has high scientific significance. For example, radial-velocity measurements with a precision of about 10 km/s can identify the position of the `snowline' in a protoplanetary disk, which is considered a key test of current models for planetary formation \cite{2016ApJ...827..113N,2017ApJ...836..118N}. The Space Infrared Telescope for Cosmology and Astrophysics (SPICA) includes the SPICA mid-infrared instrument (SMI) \cite{2017SPIE10564E..0GF,2018SPIE10698E..0CK}, which contains a high-resolution spectrometer (HR) designed to be capable of such {high-spectral-resolving-power} spectroscopy.

{In SPICA/SMI-HR, we planned to use an immersion echelle grating. 
An echelle grating is a grating that has a low groove density and is optimized for large-angle incident angles (i.e. large diffraction orders).  
In addition, in an immersion echelle grating, an echelle grating is in touch with a substrate through which the incident and diffracted beams pass. 
With a given incident angle, a large beam width incident to the grating provides us with a large spectral resolving power because it brings a large optical-path-length differences on the grating.   
The optical-path-length difference is magnified by a refractive index of a substrate in an immersion echelle grating.
Hence, a beam width required for a given spectral resolving power in an immersion echelle grating is smaller than one in usual echelle gratings by a factor of the refractive index of the substrate. 
The size reduction of the spectrometer has a particular importance in space-borne instruments that are strictly limited in size and mass\cite{2010OptEn..49e3005S}. 
}

{
As a material with a high refractive index, some semiconductors are promising in the infrared wavelength.
Crystals of Si, which transmit the light at $1.2$--$7.0\ \mathrm{\mu m}$, can be applied to immersion echelle gratings through the process of photo-lithography\cite{1993SPIE.1946..622W,2007ApOpt..46.3400M,2012SPIE.8450E..2SG}.
The single-point diamond machining\cite{2012SPIE.8442E..57S,2015ApOpt..54.5193I} of softer materials including CdZnTe is a method to fabricate immersion echelle gratings which have wavelength band pass different from Si crystals.
SPICA/SMI-HR is designed with a substrate of CdZnTe ($n$=2.65) \cite{2012SPIE.8442E..57S}.
}

{Since the size of the spectrometer is limited by cost of space craft or the available ingot size\cite{2015ApOpt..54.5193I} of the substrate material for immersion grating, it is important to obtain detailed information on the {spectral resolving power} of such a size-limited spectrometer. }
An analytical evaluation, however, does not consider either diffraction from the spectrometer slit or telescope aberrations, both of which must be taken into account to properly evaluate the {spectral resolving power}.
To include these effects, numerical simulations are needed to acquire information on the {spectral resolving power} in the MIR-wavelength band without performing difficult experiments. 
For the following reasons, the simulations must be based not on geometrical optics but on wave optics:

\begin{enumerate}
\item Diffraction from the entrance slit of a spectrometer is not negligible. Without a slit, the amplitude/phase distributions of the beam incident on the grating surface are effectively flat, as analysis with geometrical optics assumes. However, with a slit, diffraction produces beams with non-flat amplitude/phase distributions at the grating surface\cite{2022PASP..134a5002R}.
\item In the MIR, imaging performance tends to be close to the diffraction limit; therefore, it is inappropriate to use geometrical-optics-based concepts (e.g. the spot diagram radius) to estimate how the wavefront aberrations of a telescope affect the {spectral resolving power}.
\end{enumerate}

In this paper, we describe the numerical simulations of the {spectral resolving power} of the SMI--HR optical model to investigate how the presence or absence of an entrance slit affects the {spectral resolving power} of a compact, {high-spectral-resolving-power} spectrometer.
We performed the simulations using the wave-optics software Wyrowski VirtualLab Fusion (2nd Generation Technology Update [Build 7.3.1.5])\footnote{https://www.wyrowski-photonics.com/service/version-history.html}. In Section 2, we review the principle of {spectral resolving power} for immersion-echelle spectrometers to clarify the goal of the simulations. In Section 3, we describe the assumptions of our simulations. In Section 4, we show our results and discuss the general characteristics of the simulated {spectral resolving power} achieved by a compact, immersion-echelle grating spectrometer. Section 5 summarizes the contents of this paper.

\section{Review of Theory}
The theory of the {spectral resolving power} of immersion-echelle spectrometers is no different from the usual theory for ordinary reflective gratings that can be found in many sources. \cite{2000asop.conf.....S}.
We briefly review it here to clarify the aim of the simulations {(stated in Section 1)} and the symbol conventions used in this paper.
Hereafter, we assume that the light source to be observed is an ideal point source. 

The following is the fundamental equation for wavelength dispersion by a reflective grating; it is referred to as the `grating equation': 
\begin{equation}
u\left(\sin\alpha+\sin \beta \right)=\frac{m \lambda}{n},
\label{e1}
\end{equation}
where $u$ is the length of a period (the pitch) of the grating structure; $\alpha$ and $\beta$ are the angle of incidence and {diffraction} of the plane wave, respectively; $m$ is the diffraction order;  $\lambda$ is the wavelength of the light in vacuum; and $n$ is the refractive index of the {immersed grating substrate}. Although the wavelength dispersion of the medium affects the {spectral resolving power}, in this paper, we assume the refractive index\cite{1999prop.book.....B} to be a real constant, independent of $\lambda$, to focus on the other principal effects.  Note that Equation~(\ref{e1}) is derived by assuming the incident wavefront to be a perfect plane wave that spans the entire two-dimensional plane. 

 The relationship between $d\lambda$ and $d\beta$ is as follows:
 \begin{equation}
     \frac{nu\cos\beta}{m}d\beta=d\lambda.
     \label{e2}
 \end{equation}
The variation of $d\lambda/d\beta$ is negligible, compared to $d\lambda/d\beta$ itself, in the interval of the minimum resolvable exit-angle difference (denoted by $\Delta \beta$). Thus, we can interpret Equation~(\ref{e2}) as a sufficiently well-approximated relationship between $\Delta \beta$  and the minimum resolvable wavelength difference (denoted by $\Delta \lambda$):
\begin{equation}
    \frac{nu\cos\beta}{m}\Delta\beta=\Delta\lambda. 
    \label{e3}
\end{equation}
The {spectral resolving power} $R$ thus becomes: 
\begin{equation}
    R=\frac{\lambda}{\Delta\lambda}=\frac{\lambda m}{nu\cos\beta \Delta\beta}=\frac{\sin\alpha+\sin\beta}{\cos\beta \Delta\beta}.
    \label{e4}
\end{equation}
In particular, when $\alpha=\beta$ (the Littrow configuration), Equation~(\ref{e4}) becomes:  
\begin{equation}
R=\frac{2\tan \alpha}{\Delta \beta}.
\label{e5}
\end{equation}

{Hereafter, we denote the physical size of the grating window along the dispersion direction by $W$ and the geometrical beam width along the dispersion direction by $w$; and we assume that $w\leq W$.
For the development of SPICA/SMI--HR, the physical size $W$ was limited mainly by the size of the available CdZnTe ingot (e.g. a column with a diameter of 135mm and a thickness of 50mm); see \citeonline{2015ApOpt..54.5193I} for the detail of the size limitation.}

\begin{figure}[h]
     \centering
     \includegraphics[width=10cm]{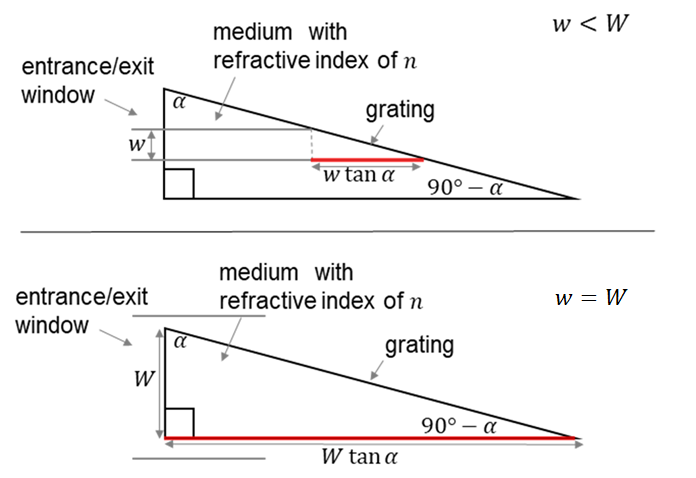}
     \caption{Illustration of the calculation of the round-trip optical-path difference between light rays from the opposite ends of the beam on the incident (exit) window. The upper panel is for the case $w<W$, and the lower panel is for $w=W$. The lengths of the solid red lines show half of the round-trip path difference between light rays from the opposite ends of the beam on the incident (exit) window. }
     \label{fig:9}
 \end{figure}

Because we are not considering diffraction from slits in this section, we can express $\Delta \beta$ as follows:
{
\begin{equation}
\Delta \beta =\eta \frac{\lambda}{nw},
\label{e7}    
\end{equation}
}
{where $\eta$ is a correction factor that depends on the shape of the entrance window; more directly, $\eta$ depends on how grating grooves are illuminated (e.g. rectangular or elliptical).}
Substituting Equation~(\ref{e7}) into Equation~(\ref{e5}) leads to the following expression for {diffraction-limited} $R$:
{
\begin{equation}
R=\frac{2n\tan \alpha w}{\eta \lambda}.
\label{e8}
\end{equation}
}
 In Equation~(\ref{e8}), we note that $R$ is directly proportional to the quantity $2n \tan \alpha w$. This quantity can be interpreted as the round-trip optical-path difference between light rays from the opposite ends of the beam width at the incident (exit) window (Figure~\ref{fig:9}).
 
{ For reference, we also consider the geometrically slit-limited case here. 
 In this case, we can express $\Delta\beta$ using the telescope diameter $D$ and the slit width in radians on the sky $\phi$ as follows:
 \begin{equation}
     \Delta \beta =\frac{1}{n}\frac{D}{w}\phi,
 \end{equation}
where $\frac{D}{w}$ equals the magnification ratio of the optics.
Hence, the geometrically slit-limited $R$ is as follows:
\begin{equation}
    R=\frac{2n\tan \alpha w}{D\phi}.
    \label{e8dash}
\end{equation}
}
 Equation~(\ref{e8}) does not consider either diffraction from the slit or telescope aberrations, both of which are needed to properly evaluate $R$. 
In other words, Equation~(\ref{e8}) is based on the assumptions that the optical system (including the telescope) is aberration-free and that there is no entrance slit.
When we consider diffraction from the entrance slit, the amplitude/phase profile incident on the grating surface becomes far more complicated than the one that analysis with geometrical optics assumes.
{In addition, diffracted beams may vignetted at the grating window.
Furthermore, aberrations are additional factors that change the size of the point-spread function (PSF).}
Spot sizes determined by ray tracing are inappropriate for evaluating the sizes of the PSFs when wavefront errors are not sufficiently large compared to the wavelength.
Hence, Equation~(\ref{e8}) is not satisfactory for evaluating the {spectral resolving power} of a realistic spectrometer with an entrance slit and/or optical aberrations. 
 
 To take these factors into account, we have performed simulations based on wave optics. The simulations assumed a simple {high-spectral-resolving-power} spectrograph of limited size ($W \approx w$). We aim to investigate how the presence or absence of an entrance slit and optical aberrations affect $R$ in the 10--20-$\mathrm{\mu}m$ wavelength range.

\section{Setup}
\subsection{Layout of the Optical Model}
\begin{figure}[htbp]
\begin{center}
\includegraphics[width=0.8\linewidth]{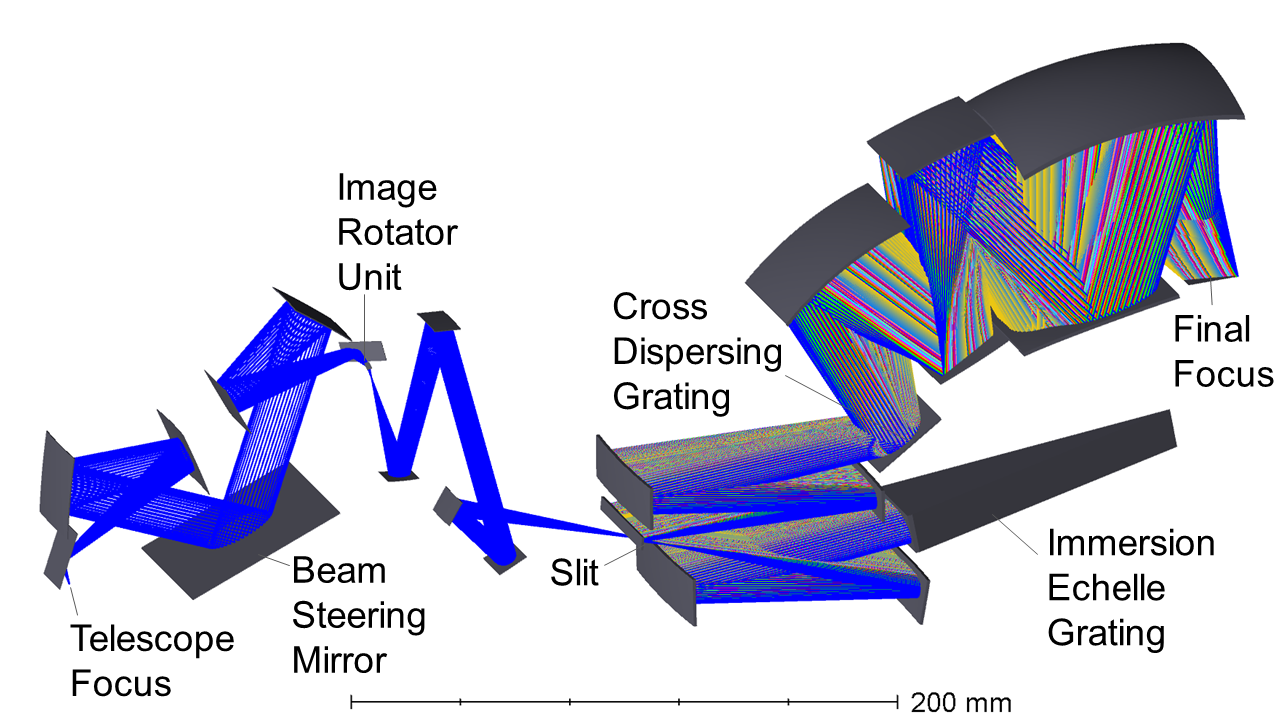} 
\end{center}
\caption{{The optical design of the SPICA/SMI-HR. The rays are colored by the diffraction orders of the diffraction by the  immersion echelle grating. The high-resolution (HR) spectrometer is one of four subunits that compose the SPICA/SMI \cite{2020SPIE11443E..6GW}.}}
\label{f11}
\end{figure}
\begin{figure}[htbp]
\begin{center}
\includegraphics[width=0.8\linewidth]{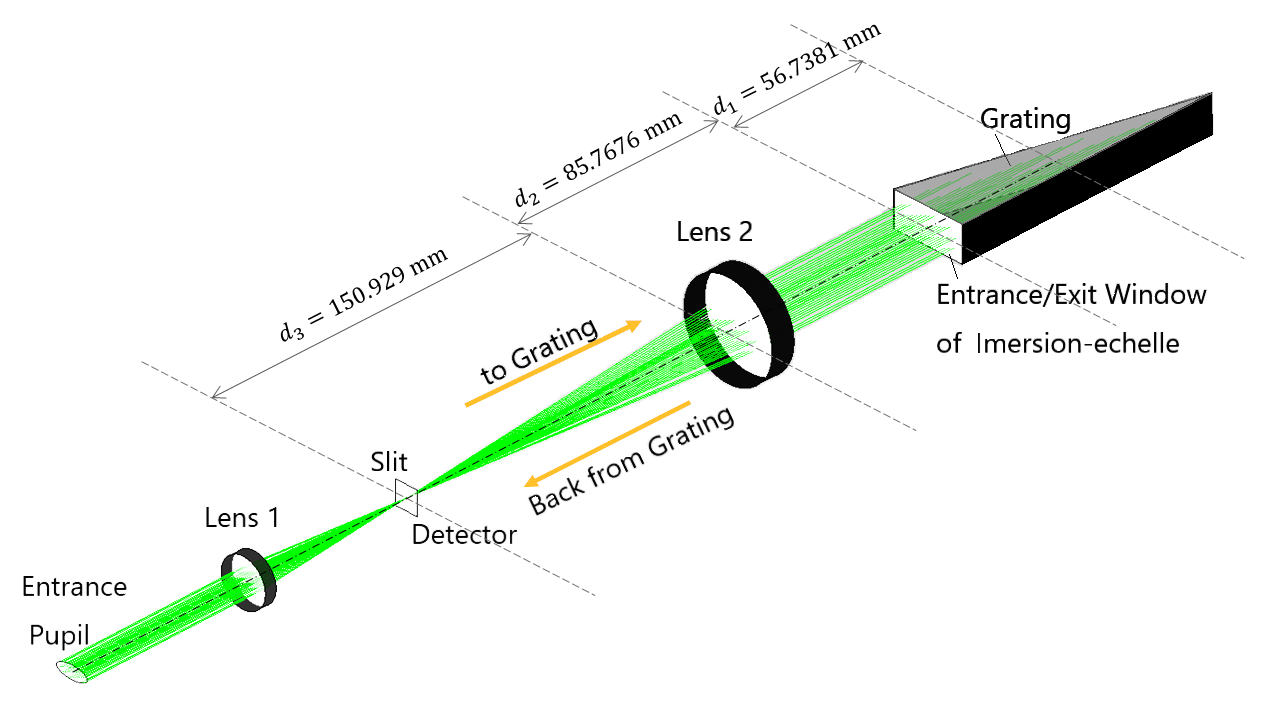} 
\end{center}
\caption{Optical layout model used for the simulations. The thicknesses of the lenses are exaggerated in this figure to make them easy to see. In the actual simulations, we used spatial phase modulations to represent thin lenses with zero thickness. {The surface-to-surface distances $d_1$, $d_2$ and $d_3$ shown here have the same definition used in Appendix B; see also Table \ref{t1}. }  }
\label{f1}
\end{figure}
\subsubsection{Simplification}
The optical model used in the current study is based on the latest design of SMI--HR\cite{2020SPIE11443E..6GW} (Figure~\ref{f11}), but we have made the following simplifications (Figure~\ref{f1}):
\begin{enumerate}
\item The cross dispersing grating, { which is used in the echelle spectrograph,} is omitted. 
\item {We simplified some focusing/collimating systems. The SPICA/SMI is designed with focusing/collimating systems consisting of multiple high-order aspherical mirrors. 
These multiple higher-order aspherical mirrors well correct off-axis aberrations. 
This compensation is necessary to obtain good imaging performance over a wide field of view for the focusing at wavelengths other than the blaze wavelength of the Littrow configuration. In the simplified model, we omitted the multiple higher-order aspherical mirrors. The omission of these items does not affect the evaluation of the spectral resolving power at the blaze wavelength of the Littrow configuration.}
\item The SMI--HR is designed with off-axis reflective optics, but our calculation model uses co-axial optics with spatial phase modulation to mimic a  thin lens of the focal length $f$. With $f$ being the focal length of the lens and $r$ representing the radial coordinate on the plane of the thin lens, we can express phase modulation by multiplication with the factor $e^{-\frac{2\pi i}{\lambda}\left(\sqrt{f^2+r^2}-f\right)}$.
We chose the spatial-phase-modulation pattern such that a spherical wave from a point source at the front focal point of the thin lens is transformed into a plane wave by the modulation.  

\end{enumerate}
\subsubsection{Main Parameters}
The main parameters of the simulation setup are compiled in Table~\ref{t1}.
They reflect the design of the SPICA/SMI--HR, except for the simplifications described above.

\begin{table}[htbp]
    \centering
    \begin{tabular}{|c|c|}
    \hline
    Parameter&Value\\\hline\hline 
Wavelength of light&$\lambda_m=\frac{2nu\sin\alpha}{m}$\\\hline
Refractive index of the immersion-echelle grating&$n=2.65000$\\\hline
Pitch of the grating&$u=282.700 \mathrm{\mu m}$\\\hline
Angle of incidence at the grating surface&$\alpha=75^{\circ}$\\\hline
Diffraction order&$m=75,85,95,105,115,125,135,$ and $145$\\\hline 
Numerical aperture (NA) in the dispersion direction&{ 0.0939085}
\\\hline
$dl/d\beta$ in Equation~(\ref{e6})& { 399.962} mm\\\hline
Entrance pupil diameter (ellipse)&14.000 mm$\times$5.60000 mm\\\hline
Distance between the entrance pupil and lens 1&{ 80.3000} mm\\\hline
Diameter of lens 1&20.0000 mm\\\hline
Thickness of lens 1&{ 0.00000 mm}\\\hline
Distance between lens 1 and slit& { 71.7112 mm}\\\hline
Aperture size of the slit&100.000 $\mathrm{\mu m}\times$720.000 $ \mathrm{\mu m}$\\\hline
Distance between the slit (or detector) and lens 2&{  150.929 mm}\\\hline
Diameter of lens 2&40.0000 mm\\\hline
Thickness of lens 2&{ 0.00000 mm}\\\hline
\begin{tabular}{c}Distance between lens 2 and the \\window of the immersion-echelle grating\end{tabular}&{ 85.7676 mm}\\\hline
Size of the window of the immersion-echelle grating&{31.5000} mm$\times$15.0000 mm\\\hline
Size of the geometrical beam at the grating window &29.4000 mm$\times$11.8000 mm\\\hline
Distance between the window and the grating center& 56.7381 mm\\\hline
Size of the grating surface&117.479 mm$\times$15.0000 mm\\\hline
\end{tabular}
\caption{The main parameters for the setup of the simulations. {These parameters come from the design of SPICA/SMI-HR. The reason why the geometry beam and grating window are elongated is to fabricate the longest possible grating from a CdZnTe ingot, which has a limited size. The width (100.000 $\mathrm{\mu m}$) of the slit are designed through our consideration of the diffraction-limited PSF width at $\lambda$=10--20 $\mathrm{\mu m}$ (Table \ref{t2}).  } }
\label{t1}
\end{table}

\subsection{Method of Calculation}
\subsubsection{Wavelength Sampling}
We performed calculations for the diffraction orders $ m = $ 75, 85, 95, 105, 115, 125, 135, and 145.
For each sampled value of $m$, we selected the following wavelengths, for which $\alpha=\beta$ (the Littrow configuration):
\begin{equation}
\lambda_m=\frac{2nu\sin \alpha}{m}.
\label{e9}
\end{equation}
These values of $\lambda_m$ are listed in Table~\ref{t2}. 
For reference, we also show the slit efficiency $E_m$ {at each wavelength, calculated for the assumed slit size $100.000 \mathrm{\mu m} \times 720.000 \mathrm{\mu m}$}.
\begin{table}[htbp]
    \centering
    \begin{tabular}{|c|c|c|c|c|c|c|c|c|}
    \hline
    $m$&145&135&125&115&105&95&85&75\\\hline
    $\lambda_m (\mathrm{\mu m})$&9.98108&10.72042&11.57805&12.58484&13.78339&15.23428&17.02654&19.29675\\\hline
    $E_m$ (\%)&87.1&86.2&84.9&83.0&80.5&77.2&72.9&67.7\\\hline
\end{tabular}
\caption{The wavelengths $\lambda_m$ used for the simulations and the slit efficiency $E_m$ calculated for the assumed slit size at each wavelength. The wavelengths $\lambda_m$ are the blaze wavelengths for  the Littrow configuration.}
\label{t2}
\end{table}
\subsubsection{Algorithms}
We next describe the propagation algorithms we used.\footnote{We used two Intel(R) Xeon(R) processors, each with CPU E5-2620 v4 (2.10 GHz, 8 cores, 16 threads), in parallel for the simulations.
The maximum amount of memory used during each simulation was less than 128 GB.} \footnote{The resolution of the simulation on each surface is 4096 $\times$ 4096.}
\begin{figure}[htbp]
\begin{center}
\includegraphics[width=0.8\linewidth]{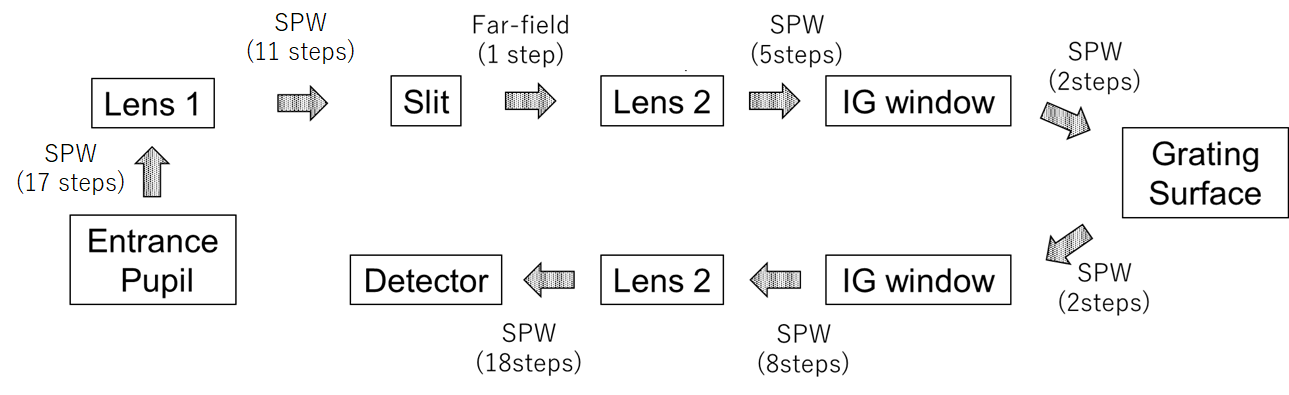} 
\end{center}
\caption{Summary illustration of the method of calculation. The abbreviation SPW represents a method that uses a spectrum of plane waves, and IG is an abbreviation for immersion-echelle grating. {We need the division of the propagation distances to reduce the amount of random-access memory that SPW requires.} However, for the slit-less cases, we calculated the propagation between lenses 1 and 2 using SPW (1 step), which is different from this figure.}
\label{f2}
\end{figure}
\begin{itemize}
    \item {A periodic dielectric media (such as a grating or lens) causes the propagated field to become periodic. Complex Fourier analysis shows that the propagated periodic field can be represented as the sum of tilted plane waves known as the Spectrum of Plane Waves (SPW).} We used the method of the SPW \cite{2005ifo..book.....G,2003JOSAA..20.1755M} to  calculate all free-space propagation, except for the interval from the slit to lens 2 (Figure~\ref{f2}).
This exception is necessary because, with SPW, a large amount of memory is required to calculate large-angle diffraction after the slit.
The SPW method also requires a large amount of memory to calculate the propagation between planes separated by a large distance.
Hence, we split each interval between the optical elements into several smaller calculation steps. 
The number of such divisions is indicated in parentheses in Figure~\ref{f2}.
\item We used the far-field approximation to Rayleigh's diffraction formula of the first kind \cite{1999prop.book.....B} for the propagation from the slit to lens 2 (Figure~\ref{f2}), {which requires a huge amount of memory for large angle diffraction in the case of the SPW method.} The operator $\mathcal{P}^{\mathrm{Far Field}}_{\Delta z}$ for the far-field approximation is as follows: 
\end{itemize}
\begin{equation}
[\mathcal{P}^{\mathrm{Far Field}}_{\Delta z}U(x,y)](x',y')=\frac{\Delta z}{i\lambda r'}\frac{e^{\frac{2\pi i r'}{\lambda}}}{ r'}\int_{-\infty}^{\infty}\int_{-\infty}^{\infty} \! \! dx dy U(x,y) e^{\frac{2\pi i (x'x+y'y)}{\lambda r}}, 
\label{e10}    
\end{equation}
where $(x,y)$ and $(x',y')$ are Cartesian coordinate systems on the surfaces before and after the propagation, $U(x,y)$ is the original field before the propagation, $\Delta z$ is the distance between the two planes along the optical axis, and  
\begin{eqnarray}
r&=&\sqrt{x^2+y^2+(\Delta z)^2} \nonumber \\
r'&=&\sqrt{x'^2+y'^2+(\Delta z)^2}. 
\label{e11}    
\end{eqnarray}

\begin{itemize}
\item We modeled the grating as a linear phase-modulation pattern. This model can be interpreted as a grating that has an ideally selective diffraction efficiency; in other words, the diffraction efficiency of this grating model is 100\% for an observed diffraction order and 0\% for the other orders; see Appendix~A for details. 
\item We used Equation~(\ref{e5}) to evaluate $R$. We evaluated  $\Delta\beta$ in Equation~(\ref{e5}) using the width of the calculated monochromatic PSF, which we denote by $\Delta l$, through the following relationship:
\end{itemize}
\begin{equation}
\Delta\beta=\left(\frac{dl}{d\beta}\right)^{-1} \Delta l.
\label{e6}
\end{equation}
The value of $dl/d\beta$ in Equation~(\ref{e6}) depends on the optical design (see Table~\ref{t1} and Appendix~B).
 Using the value of $dl/d\beta$ given in Table~\ref{t1}, together with Equation~(\ref{e5}) and (\ref{e6}), we obtain the following expression:
\begin{equation}
R=\frac{2985356\mathrm{\mu m}}{\Delta l}.
\label{e100}
\end{equation}
The definition of $\Delta l$ \cite{2013PASA...30...48R} used in this paper is the full width at half maximum (FWHM) along the direction of wavelength dispersion; we first integrated the PSFs along the direction perpendicular to the direction of wavelength dispersion and then used them to evaluate the FWHMs.

\section{Results and Discussion}
\subsection{Results}

\begin{figure}[htbp]
\begin{center}
\includegraphics[width=0.8\linewidth]{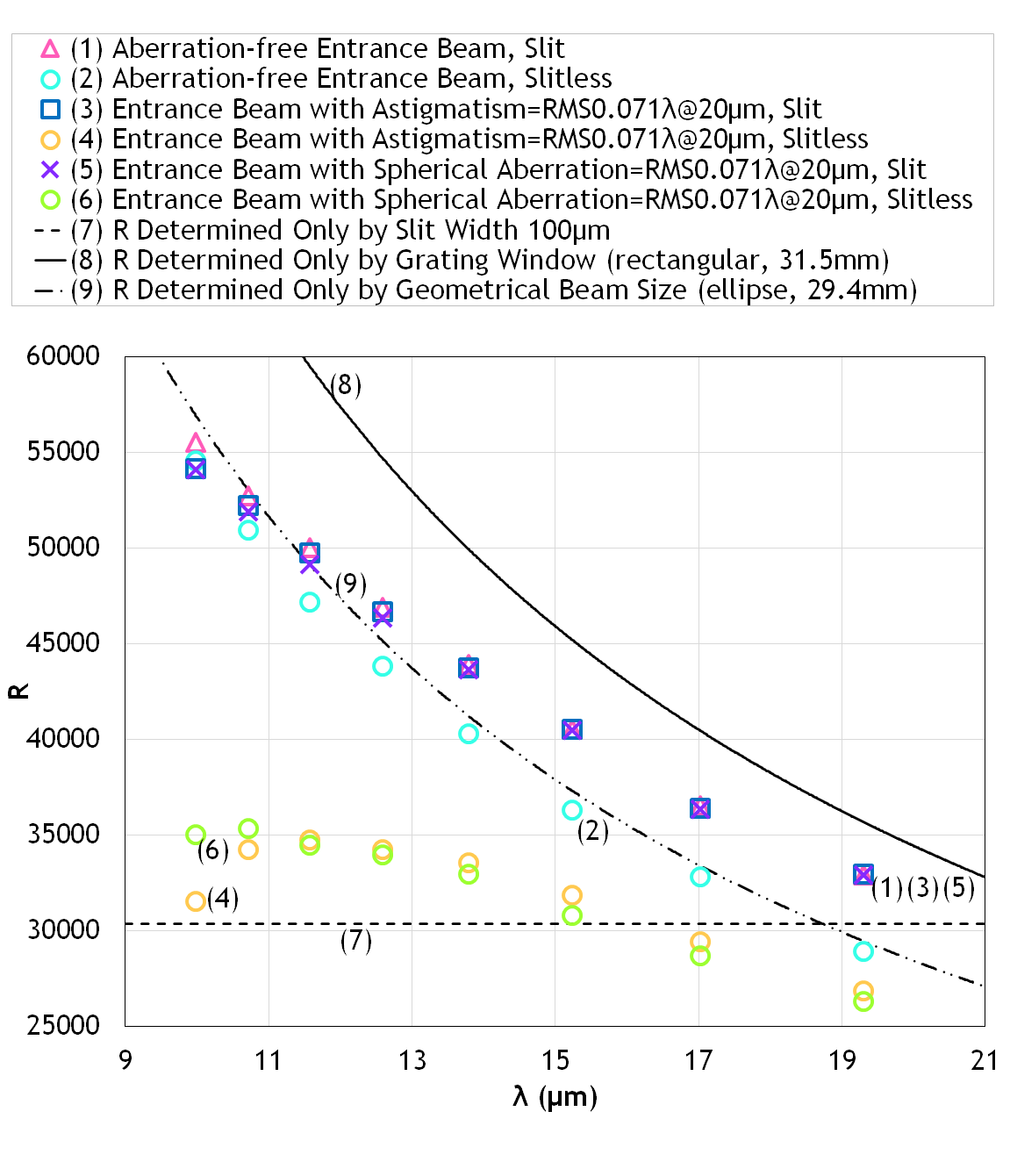} 
\end{center}
\caption{{ The simulated {spectral resolving power} at different wavelengths. {The symbols indicate the results of numerical simulations with SPW modeling, while the curves indicate the predictions of the analytical calculations.} The magenta triangles (1) show simulated {spectral resolving power} for an  aberration-free case with a 100-$\mathrm{\mu m}$ slit. The blue squares (3) and purple crosses (5) represent the results for the cases with a  100-$\mathrm{\mu m}$ slit, but they include astigmatism and spherical aberration, respectively. In each case, the magnitude of the assumed wavefront aberration is RMS $0.071\lambda_{\mathrm{ref}}$ (where  $\lambda_{\mathrm{ref}}=20 \mathrm{\mu m}$). The circles indicate the results of the slit-less cases; the cyan (2), orange (4) and green (6) circles represent the aberration-free results, the results with astigmatism, and the results with spherical aberration, respectively. The assumed aberration is the same as that in the cases with a slit. The three black curves (7--9) are the results of some analytical estimates (see the Discussion section in the main text).
}}
\label{f3}
\end{figure}

The simulated {spectral resolving power}s for different wavelengths are shown in Figure~\ref{f3}; for comparison, the results for slit-less cases are also shown. 
For cases both with and without a slit, we include the following three types of aberrations {(added as wavefront errors at the entrance pupil of the simulation model)}: aberration-free [(1) and (2) in Figure~\ref{f3})], astigmatism with RMS $0.071\lambda_{\mathrm{ref}}$ [(3) and (4) in Figure~\ref{f3}] and spherical aberration with RMS $0.071\lambda_{\mathrm{ref}}$ [(5) and (6) in Figure~\ref{f3}]. The symbols (1), (3) and (5) are for cases with a slit, while (2), (4) and (6) are for cases without a slit.

\begin{figure}[htbp]
\begin{center}
\includegraphics[width=0.8\linewidth]{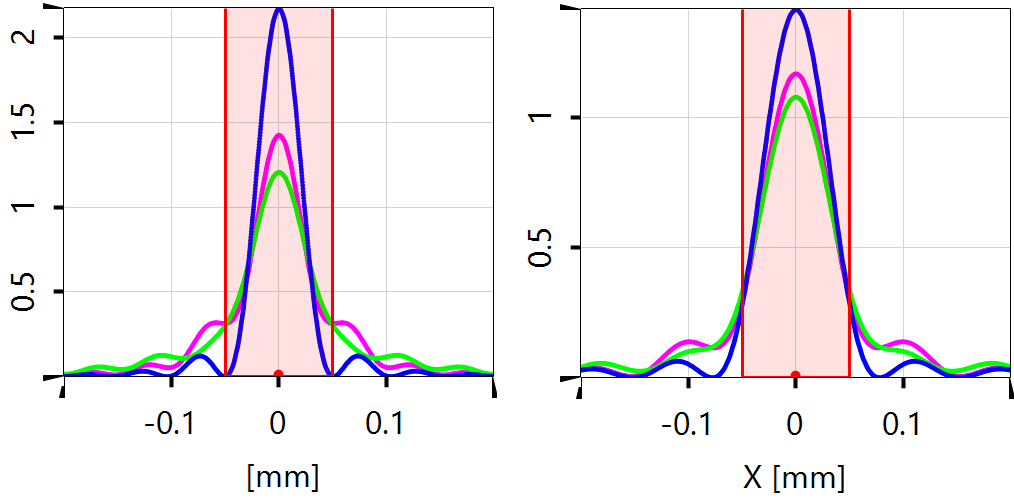} 
\end{center}
\caption{{The one-dimensional PSFs on the slit plane at $9.98108\ \mathrm{\mu m}$ (left) and $15.23428$ $\mathrm{\mu m}$ (right) of the wavelength. The vertical axis is the relative intensity. The blue, magenta, green curves indicate the cases of aberration-free, astigmatism and spherical aberration, respectively. The red filled area indicates the width of the slit aperture.}}
\label{fSLIT}
\end{figure}

\begin{figure}[htbp]
\begin{center}
\includegraphics[width=0.8\linewidth]{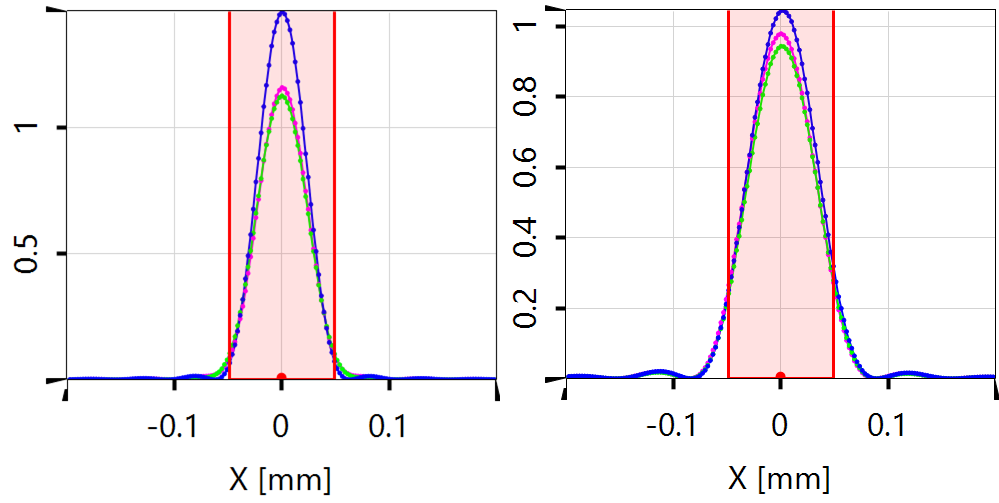} 
\end{center}
\caption{{The one-dimensional PSFs on the final focal plane at $9.98108\ \mathrm{\mu m}$ (left) and $15.23428$ $\mathrm{\mu m}$ (right) of the wavelength. The vertical axis is the relative intensity. The blue, magenta, green curves indicate the cases of aberration-free, astigmatism and spherical aberration, respectively. The red filled area indicates the width of the slit aperture.}}
\label{fFF}
\end{figure}

{For reference, Figure \ref{fSLIT} shows the one-dimensional PSFs on the slit plane at the blaze wavelength of the diffraction orders $m=145, 95$ (Table $\ref{t2}$); in addition, Figure \ref{fFF} shows the one-dimensional PSFs on the final focal plane at the same wavelengths as Figure \ref{fSLIT}.
We obtained these one-dimensional PSFs through integrating the two-dimensional amplitude distribution along the direction perpendicular to the direction of wavelength dispersion.}

\subsection{Discussion}
In Figure~\ref{f3}, the three curves [(7--9) in Figure~\ref{f3}] represent some simple analytical estimates of $R$. In this section, we first explain those three curves and then discuss the results using them.
In the discussion below, the distance between lens 2 and the detector is denoted by $f$ ($=150.929\ \mathrm{mm}$).
We define two types of numerical apertures; one ($\mathrm{NA}_{GW}$) is associated with the size $W$ of the grating window along the dispersion direction, and the  other ($\mathrm{NA}$) is for the geometrical beam size $w$ along the dispersion direction:
\begin{eqnarray}
 \mathrm{NA}_{GW}&=&\sin{\left(\arctan{\left(\frac{W}{2f}\right)}\right)} \nonumber \\
 \mathrm{NA}&=&\sin{\left(\arctan{\left(\frac{w}{2f}\right)}\right)}.
\end{eqnarray}
The values of $\mathrm{NA}_{GW}$ and $\mathrm{NA}$ are 0.100550 and 0.0939085, respectively.

 The horizontal dashed black line (7) represents the value of $R$ determined only by the slit width, i.e. the value of $R$ for the case in which $\Delta l$ is the  full slit width (100 $\mathrm{\mu m}$).
 {This line corresponds to a specific example of Equation (\ref{e8dash}).}
 This line provides a good estimate of $R$ for the sufficiently diffuse sources in all the wavelength ranges discussed in the present paper. \footnote{Exceptions are extremely long wavelengths for which we cannot use Kirchhoff's boundary conditions.\cite{1999prop.book.....B}.}
  Next, the black solid curve (8) and the dash-dot-dot curve (9) represent $R$ for cases in which $\Delta l$ is evaluated from the following equations, respectively: 
\begin{equation}
\Delta l=\Upsilon_{\mathrm{rect}}\lambda/(2\mathrm{NA}_{GW})
\label{EQ}
\end{equation}
and
\begin{equation}
\Delta l=\Upsilon_{\mathrm{elli}}\lambda/(2\mathrm{NA}),
\label{EQEQ}
\end{equation}
where $\Upsilon_{\mathrm{rect}}$ ($=0.8859$) and $\Upsilon_{\mathrm{elli}}$ ($=1.002$).
 {These lines correspond to specific examples of Equation (\ref{e8}).}
{Equations (\ref{EQ}) and (\ref{EQEQ}) are obtained with the calculation of the Fraunhofer diffraction integral for rectangular and elliptical aperture functions \cite{1999prop.book.....B}; these expressions are not used in the SPW simulations.}

Curve (8) gives the value of $R$ for the case in which the grating window is fully and homogeneously illuminated by a perfect plane wave; thus, this curve can be interpreted as the upper limit on $R$, which is  realized approximately for a case with a sufficiently narrow slit width.
{In other words, when the collimated beam projection is less than the physical extent of the grating ($w<W$), inserting a narrow slit spreads the beam over the entire grating, reproducing the result where $w=W$ without a slit.}
Similarly, curve (9) shows the value of $R$ for a case in which the geometrical beam size determines the illuminated region on the grating window.
When we compare the symbols (1--6) and the lines (7--9) in Figure~\ref{f3}, we find the followings:
\begin{enumerate}[(i)]
\setcounter{enumi}{0}
    \item Symbols (2) fall approximately on curve (9).
\end{enumerate}
This means that the numerical and analytical estimates of $\Delta l$ are approximately consistent with each other. 
\begin{enumerate}[(i)]
\setcounter{enumi}{1}
    \item Symbols (1), (3) and (5) trace almost the same positions.
    \end{enumerate}
This can be interpreted as meaning that the slit suppresses the effect of the assumed aberrations on $R$.
This is significantly different from the previous estimate of the SPICA SMI--HR {spectral resolving power} \cite{2020SPIE11443E..6GW}.
The previous estimate is based on geometrical optics\footnote{Wada et al. (2020)\cite{2020SPIE11443E..6GW}  estimated the {spectral resolving power} of SMI--HR assuming a telescope-PSF size diffraction-limited at 20$\mathrm{\mu m}$ and geometrical optics for SMI--HR.}  and R takes almost the same values as line (7).
Experimental demonstration of the improvement of the {spectral resolving power} R for the cases with aberrations by the existence of a slit would be valuable.
\begin{enumerate}[(i)]
\setcounter{enumi}{2}
    \item Symbols (1), (3) and (5) lie in the region between curves (8) and (9) for $\lambda\gtrsim 12 \mathrm{\mu m}$.
    \end{enumerate}
This is a result of diffraction from the slit spreading the illuminated region of the grating surface. 
Since the assumed width of the slit is not sufficiently larger than the widths of diffraction-limit PSFs, an electromagnetic wave that has passed through the slit has a different angular spectrum compared to the wave before passing through the slit. 
Hence, electromagnetic waves that have passed through the slit illuminate the grating surface in different ways from the assumption of curve (8).

\begin{figure}[htbp]
\begin{center}
\includegraphics[width=0.8\linewidth]{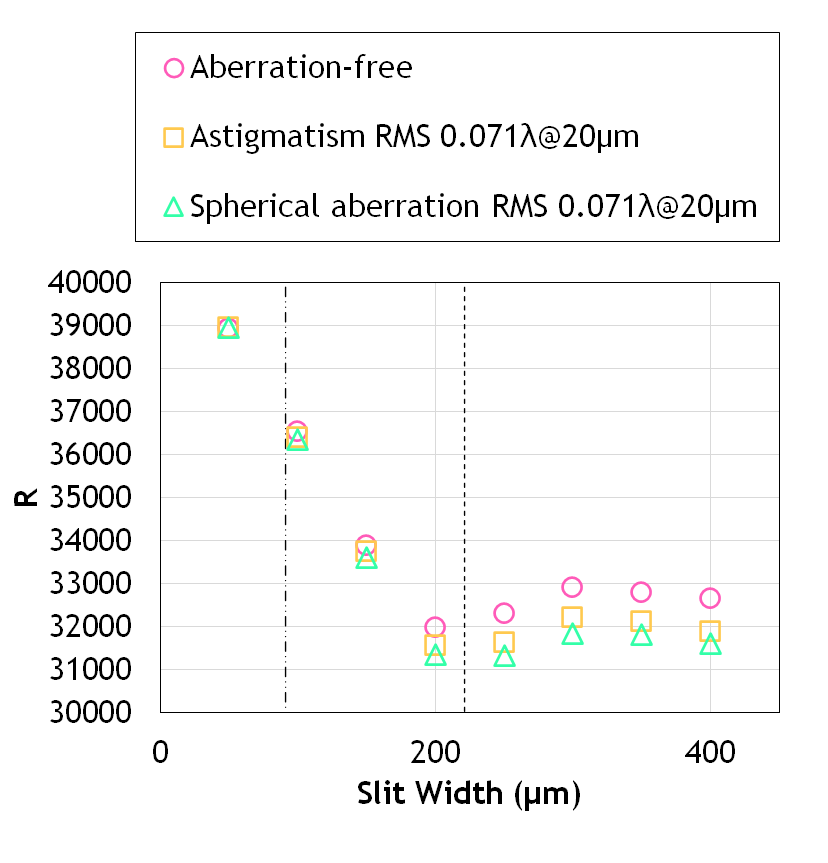} 
\end{center}
\caption{{Spectral resolving power} in cases with different slit widths ($m=85,\ \lambda=17.02654\ \mathrm{\mu m}$).  The magenta circles show the simulated {spectral resolving power} for the aberration-free case. The orange squares and green triangles represent the results for the cases with astigmatism and with spherical aberration, respectively. For each case, the magnitude of assumed wavefront aberration is RMS $0.071\lambda_{\mathrm{ref}}$, where $\lambda_{\mathrm{ref}}=20 \mathrm{\mu m}$. The black dash-dot-dot line represents the FWHM (for the short-side direction of the slit) of the PSF incident on the slit. The first-null point along the axis of the short-side direction of the slit (in the aberration-free case) is located at a half width of 110.60 $\mu m$; that is, at a full width of 221.20 $\mu m$. The 221.20-$\mu m$ width is  shown as a vertical dashed black line.}
\label{f4}
\end{figure}

Figure~\ref{f4} shows the results for the simulations of the {spectral resolving power} for cases with slit widths of 50, 100, 150, 200, 250, 300, 350 and 400 $\mathrm{\mu}$m ($m=85,\ \lambda=17.02654\ \mathrm{\mu m}$).
For each slit width, we examined three types of aberrations (aberration-free; astigmatism with RMS $0.071\lambda_{\mathrm{ref}}$, where $\lambda_{\mathrm{ref}}=20 \mathrm{\mu m}$; and spherical aberration with RMS $0.071\lambda_{\mathrm{ref}}$). Hereafter, we express the full width of the first-null point along the axis of the short-side direction of the slit (221.20 $\mu m$) as $\epsilon$.

Figure~\ref{f4} shows the following tendencies:
\begin{enumerate}
\item For all the aberrations, in the region where the slit width is narrower than $\epsilon$, $R$ declines approximately linearly as the slit width is increased.
This indicates that narrow slits enhance the {spectral resolving power} even if the window of the immersion echelle grating is as compact as the geometrical beam width.
\item As the slit width increased beyond $\epsilon$, $R$ increases again after first passing through a local minimum. To interpret this phenomenon, it is useful to consider the one-dimensional Fraunhofer diffraction model for the pupil function: 
\begin{equation}
    \Pi(q)= \left\lbrace\begin{array}{cc}
        0 & \left(|q|>\frac{1}{2}\right) \\
        \frac{1}{2} & \left(|q|=\frac{1}{2}\right) \\
        1 & \left(|q|<\frac{1}{2}\right)
    \end{array}
    \right.
    \label{A16}
    ;
\end{equation}
that is 
\begin{equation}
     \int_{-\infty}^{\infty}dq\Pi(q)e^{-2\pi i \xi q} = \frac{\sin{(\pi \xi)}}{\pi \xi},
     \label{A17}
\end{equation}
where $q$ is  the pupil coordinate normalized by the pupil diameter and $\xi$ is the focal-plane coordinate normalized by  $\lambda/(2\mathrm{NA})$ ($=\epsilon/2$). Here, we assume that the slit aperture is $\Pi\left(\xi/p\right)$, where $p$ is a parameter that represents the width of the slit (the full-width of the slit normalized by $\epsilon/2$). Then, by assuming that the inverse Fourier transform of $\Pi\left(\xi/p\right)\sin{(\pi \xi)}/\left(\pi \xi\right)$ gives the amplitude on the grating surface, we can write that amplitude as follows (the finite size of the grating is ignored here):
\begin{equation}
     \int_{-\infty}^{\infty}d\xi\Pi\left(\frac{\xi}{p}\right)\frac{\sin{(\pi \xi)}}{\pi \xi}e^{2\pi i \xi q}=\frac{\mathrm{Si}\left(p \pi\left(\frac{1}{2}+q \right)\right)+\mathrm{Si}\left(p \pi\left(\frac{1}{2}-q \right)\right)}{\pi},
     \label{e36}
\end{equation}
where $\mathrm{Si}(z)$ is the sine integral:
\begin{equation}
    \mathrm{Si}(z)=\int _{0}^{z}dt\frac{\sin t}{t}.
    \label{A19}
\end{equation}
\begin{figure}[htbp]
\begin{center}
\includegraphics[width=1.0\linewidth]{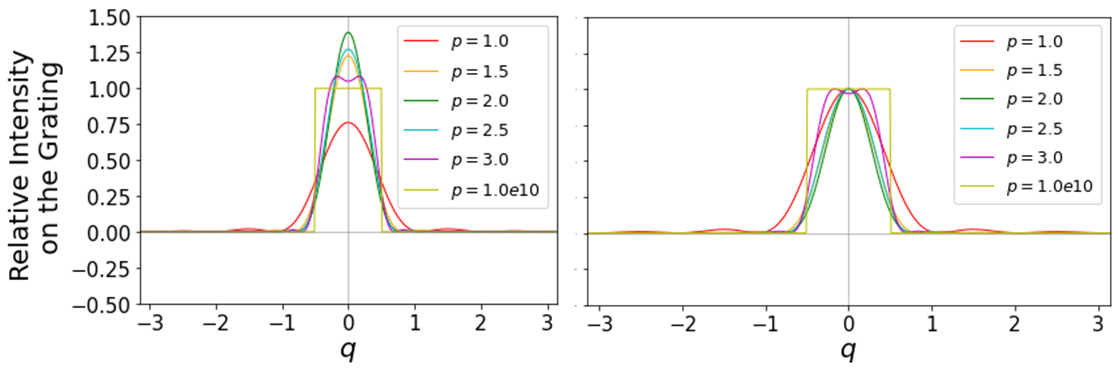} 
\end{center}
\caption{{Left:} the functional form of the square of the right-hand side of Equation~(\ref{e36}) for different values of $p$; in other words, the intensity profiles on the grating surface for different slit widths (for the one-dimensional model described in the main text). The horizontal axis is the coordinate on the grating surface normalized by the full width of the grating window. The vertical axis is the relative intensity of light incident on the grating. The parameter $p$ is the full-width of the slit normalized by $\epsilon/2$. The curve for the case $p=1.0\times10^{10}$ is shown to represent the slit-less case. {Right: the same curves as the left panel that is normalized by their own peak values for clarity of their peak widths.} }
\label{f6}
\end{figure}
Figure~\ref{f6} shows the functional forms of the right-hand side of Equation~(\ref{e36}) for different values of $p$; in other words, these are the amplitude profiles on the grating surface for different slit widths; the local minimum in Figure~\ref{f4} corresponds approximately  to the case $p=2$. The curve for the case $p=1.0\times10^{10}$ is shown to represent the slit-less case; i.e. this curve shows the amplitude profile of the original pupil aperture. Figure~\ref{f6} shows  that the amplitude of the case $p=2.0$ is concentrated in a small region on the original pupil aperture, compared with the other cases. As $p$ becomes smaller than 2.0, diffraction from the slit enlarges the effective illuminated regions, and $R$ is thus increased. Conversely, as $p$ becomes greater than 2.0, the amplitude profiles change from bell-shaped curves to top-hat-like shapes. This enlarges the effective illuminated region on the grating until it is as large as the original pupil aperture. Hence, $R$ is increased in this case as well.
\item The difference amongst the three cases of aberration is not obvious in the region where the slit width is narrower than $\epsilon$, while the difference is obvious when the slit is wider than $\epsilon$.
This occurs because a narrow slit works as a spatial filter that suppresses the impact of aberrations.
\end{enumerate}

\begin{enumerate}[(i)]
\setcounter{enumi}{3}
    \item Symbols (1), (3), and (5) lie approximately on curve (9)  for $\lambda\lesssim 12 \mathrm{\mu m}$.
    \end{enumerate}
Since the sizes along the dispersion direction of beams incident on the slit are proportional to the wavelengths for the case without optical aberration, the extent of the slit-diffraction effect at short wavelengths is expected to be smaller than that at longer wavelengths. 
\begin{enumerate}[(i)]
\setcounter{enumi}{4}
    \item Symbols (4) and (6) have {spectral resolving power}s lower than symbols (2). 
\end{enumerate}
This can be interpreted as the effect of the assumed aberration.
Here we estimate the effect of the assumed aberration on $R$ as follows.
The {typical} intensity $\left<I\right>$ of the PSF is defined as follows: 
\begin{equation}
    \left<I\right> = \frac{P}{s(\Delta l)^2},
    \label{e17}
\end{equation}
where $P$ is the power at the last focal plane (the two-dimensional integral of the intensity over the whole of the last focal plane) and $s$ is a dimensionless constant, the actual value of which is arbitrary and does not affect the result. Since wavefront aberrations do not change $P$, Equation~(\ref{e17}) shows that the value of $\left<I\right>(\Delta l)^2$ is constant, independent of wavefront aberrations. Therefore, by assuming as a working hypothesis that $ \left<I\right> $ is proportional to the Strehl ratio associated with the considered aberration {(i.e. an assumption that the Strehl ratio is proportional to the typical PSF intensity $\left<I\right>$)}, we obtain the following relation:  
\begin{equation}
     \Delta l \propto \mathrm{(Strehl\ ratio)}^{-\frac{1}{2}}.
\end{equation}
Thus, since $R$ is inversely proportional to $\Delta l$ [Equation~(\ref{e5}) and (\ref{e6})], $R$ (for the slit-less cases) is proportional to the square root of the Strehl ratio. 

\begin{figure}[htbp]
\begin{center}
\includegraphics[width=0.8\linewidth]{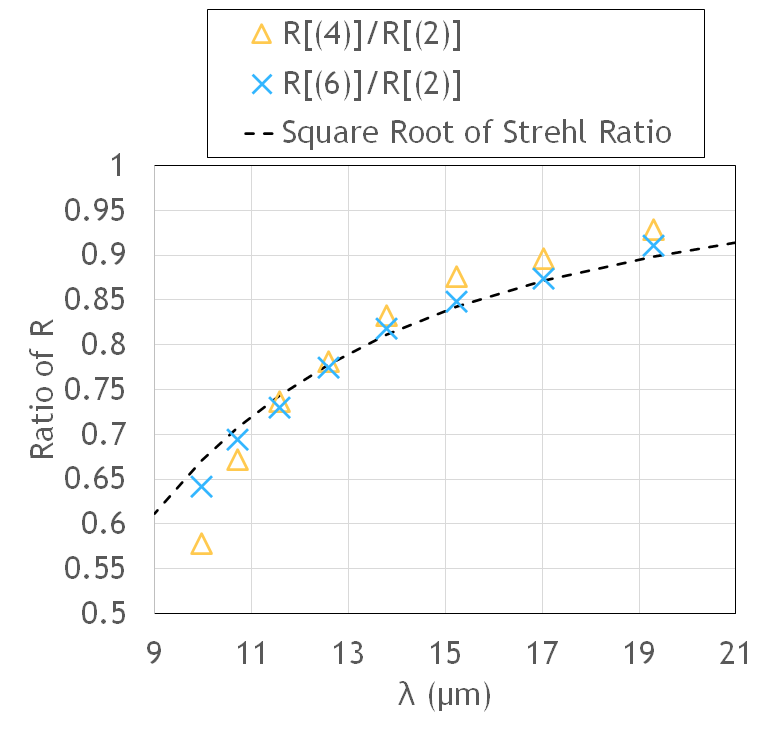} 
\end{center}
\caption{The impact of aberrations on $R$ in the cases without a slit. The numbers in the legend denote the model identification numbers in Figure~\ref{f3}. The orange triangles represent the ratio of $R$ for a  case with astigmatism [(4) in Figure~\ref{f3}] to the value of $R$ for a case without aberration [(2) in Figure~\ref{f3}]. Similarly, the blue crosses represent the ratio of $R$ for a case with spherical aberration [(6) in Figure~\ref{f3}] to the value of $R$ for a case without aberration [(2) in Figure~\ref{f3}]. The black dashed curve shows the square root of the Strehl ratio associated with the considered aberrations.}
\label{f5}
\end{figure}
Figure~\ref{f5} shows $R$ for cases (4) and (6) in Figure~\ref{f3} divided by $R$ for case (2), together with the square root of the Strehl ratio; note that the Strehl ratio for case (2) is approximately unity {(diffraction limited)}. 
Here, we define the Strehl ratio as follows:
\begin{equation}
    \mathrm{(Strehl\ ratio)}=e^{-\left(2\pi \Delta q \right)^2},
\end{equation}
where $\Delta q$ is the RMS wavefront error {in waves}.
Figure~\ref{f5} shows that the square root of the Strehl ratio is valid as a rough estimate of the impact of aberrations on the value of $R$ in the cases without a slit.

\section{Conclusion}
The commonly-used theory for the {spectral resolving power} of an immersion-echelle grating does not take into account diffraction from the slit, optical aberration, or two-dimensional aperture shapes and beam profiles. 
In this study, we performed numerical simulations to take these factors into account to properly investigate the {spectral resolving power} of a compact, {high-spectral-resolving-power} MIR spectrometer.
This simulation was fully based on wave optics, as computed using  Wyrowski VirtualLab Fusion (2nd Generation Technology Update [Build 7.3.1.5]).

Our main results are as follows:
\begin{itemize}
  \item Because diffraction from the slit spreads the illuminated region of the grating surface, when there are no optical aberrations, $R$ is larger for cases with a slit than for cases without a slit.
  \item {The spectral resolving power $R$ reaches a minimum when the slit width is roughly at the width between the aberration-free-PSF's first null points along the axis of the short-side direction of the slit.  As the slit widens more, $R$ increases again slightly thanks to a more uniform illumination of the grating.}
    \item For cases with a slit (with the slit width assumed to be 94\% of the FWHM of the aberration-free PSF at 20 $\mathrm{\mu m}$), the impact of the assumed aberrations on $R$ is small and can be ignored {thanks to the spatial-filtering effect by the slit.} This result is completely different from the prediction of geometrical optics, where $R$ depends only on the assumed aberration and is independent of $\lambda$. {
In practical terms, it is important to consider the trade-off relationship between spatial filtering and a throughput since a narrower slit reduces the light throughput especially for aberrated PSFs.}
      \item The slit-diffraction effect on $R$ is weaker at the short-wavelength region than at the long-wavelength region.
       \item For cases without a slit, the impact of aberrations on $R$ can be roughly estimated using the square root of the Strehl ratio associated with the considered aberration.    
\end{itemize}

\section*{Acknowledgments}
This research is part of conceptual design activity for the infrared astronomical space mission SPICA, which was a candidate for the ESA Cosmic Vision M5 and a JAXA strategic L-class mission.
The SMI consortium that is one of the authors of this paper is an international team who was in charge of the development of SPICA Mid-Infrared Instrument; this consortium is led by Japanese universities and ISAS/JAXA.
\section*{Appendices}
\section*{A\hspace{10px}Model of a Reflective Diffraction Grating with Ideal Diffraction Efficiency}
The x--y plane on the grating surface is defined as follows:
\begin{itemize}
    \item x is the coordinate perpendicular to the grooves on the grating surface.
    \item y is the coordinate parallel to the grooves on the grating surface.
\end{itemize}


For simplicity, assume that the diffraction grating 
is infinitely long in the y-direction. Also, assume that the diffraction grating modulates the incident plane wave on the grating surface as follows: 
\begin{eqnarray}
 g(x)\mathrm{rect}(x/D), 
 \label{A1}
\end{eqnarray}
where    
\begin{equation}
    g(x+u)=g(x),
\label{A2}
\end{equation}
$u$ and $D$ are the pitch and full width of the grating, and
\begin{equation}
\mathrm{rect}(x)= 1\ \ (|x|<0.5),\ 0.5\ \  (|x|=0.5),\ 0\ \  (|x|>0.5).  
\label{A3}
\end{equation}
Because $g(x)$ is a periodic function with period $u$, the Fourier conjugate \footnote{This Fourier transform can be physically interpreted as a plane-wave expansion in three-dimensional space.} of $g(x)\mathrm{rect}(x/D)$ is as follows:
\begin{equation}
    \mathcal{F}[g(x)\mathrm{rect}(x/D)](n\kappa_x)=\left(S(n\kappa_x)\sum_{m=-\infty}^\infty\delta(n\kappa_x-m/u)\right)\ast \frac{\sin(D\pi n\kappa_x)}{\pi n\kappa_x},
    \label{A4}
\end{equation}
where $m$, $n$ and $S(n\kappa_x)$ are the diffraction order, index of refraction, and diffraction efficiency, respectively, and where  $\kappa_x=k_x/(2\pi)$ and $k_x$ is the wavenumber in the x-direction. 
Here, we assume an ideal diffraction efficiency; that is
\begin{equation}
S(n\kappa_x)\sum_{m=-\infty}^\infty\delta(n\kappa_x-m/u)=\delta(n\kappa_x-m_{\mathrm{design}}/u),
\label{A5}
\end{equation}
where $m_{\mathrm{design}}$ means the diffraction order intended to be observed. Equation~(\ref{A4}) then becomes:
\begin{equation}
    \mathcal{F}[g(x)\mathrm{rect}(x/D)]=\frac{\sin(D\pi(n\kappa_x-m_{\mathrm{design}}/u))}{\pi(n\kappa_x-m_{\mathrm{design}}/u)}.
    \label{A6}
\end{equation}
The inverse Fourier transform of Equation~(\ref{A6}) is as follows:
\begin{equation}
    g(x)\mathrm{rect}(x/D)=e^{2\pi i m_{\mathrm{design}}x/u }\mathrm{rect}(x/D).
    \label{A7}
\end{equation}
On the other hand, the grating equation for a reflective grating is as follows:
\begin{equation}
u(\sin\alpha+\sin\beta)=m_{\mathrm{design}}\lambda/n.
\label{A8}
\end{equation}
Using $n\kappa_x$, Equation~(\ref{A8}) can be rewritten in the following form:
\begin{equation}
n\kappa_{x0}+n\kappa_x=m_{\mathrm{design}}/u,
\label{A9}
\end{equation}
where $\kappa_{x0}$ is $\sin(\alpha)/\lambda$. By substituting Equation~(\ref{A9}) into Equation~(\ref{A7}), we obtain the following:
\begin{equation}
    g(x)\mathrm{rect}(x/D)=e^{2\pi i n (\kappa_{x0} x+\kappa_x x) }\mathrm{rect}(x/D).
    \label{A10}
\end{equation}
Equation (\ref{A10}) shows that, when we assume an ideal diffraction efficiency [Equation~(\ref{A5})], we can model a diffraction grating as a `mirror'  that reflects plane waves from the direction of $\kappa_{x0}$ to the direction of $\kappa_x$. In the simulations, we define the direction of $\kappa_{x0}$ as the peak direction of the spectrum of the plane wave incident on the grating.

\section*{B\hspace{10px}Evaluation of $dl/d\beta$\label{ApB}}
{ The quantity $dl/d\beta$ can be evaluated using matrix ray-tracing based on Gaussian optics \cite{hecht2002optics}. We explain the method here. Hereafter, we denote the distance from the center of the grating surface to the grating window by $d_1$, the distance from the grating window to the thin lens by $d_2$ and the distance from the thin lens to the detector by $d_3$.

The ray height, $h$, and the angular direction of the ray, $\theta$, in radians can be written in the following vector form:
\begin{equation}
\begin{pmatrix}
h \\
\theta \\
\end{pmatrix}.
\label{A11}
\end{equation}
Thus, the vector:
\begin{equation}
\begin{pmatrix}
0 \\
d\beta \\
\end{pmatrix}
\label{A12}
\end{equation}
means the chief ray (on the grating surface) with the tiny first-order-direction angle $d\beta$. 
 
The transfer matrix from the grating surface to the detector can then  be calculated as follows:  
\begin{equation}
\begin{pmatrix}
1&d_3 \\
0&1 \\
\end{pmatrix}
\begin{pmatrix}
1&0 \\
-\frac{1}{d_3}&1 \\
\end{pmatrix}
\begin{pmatrix}
1&d_2 \\
0&1 \\
\end{pmatrix}
\begin{pmatrix}
1&0 \\
0&n \\
\end{pmatrix}
\begin{pmatrix}
1&d_1 \\
0&1 \\
\end{pmatrix}
= 
\begin{pmatrix}
0&n d_3 \\
-\frac{1}{d_3}&n\left(1-\frac{d_2}{d_3}\right)-\frac{d_1}{d_3} \\
\end{pmatrix}
.
\label{A13}
\end{equation}

Thus, the chief ray with the tiny direction angle on the grating surface propagates onto the detector as given by the following equation:
\begin{equation}
\begin{pmatrix}
0&n d_3 \\
-\frac{1}{d_3}&n\left(1-\frac{d_2}{d_3}\right)-\frac{d_1}{d_3} \\
\end{pmatrix}
\begin{pmatrix}
0 \\
d\beta \\
\end{pmatrix}
=
\begin{pmatrix}
n d_3 d\beta\\
\left(n\left(1-\frac{d_2}{d_3}\right)-\frac{d_1}{d_3}\right)d\beta \\
\end{pmatrix}
.
\label{A14}
\end{equation}

Since the height of the ray on the detector can be interpreted as $dl$, we obtain the following expression for $dl/d\beta$:
\begin{equation}
    dl/d\beta=n d_3.
    \label{A15}
\end{equation}
}

\bibliographystyle{spiejour}
\bibliography{sample}

\end{document}